\documentclass[a4paper,fleqn,usenatbib]{mnras}

\usepackage[T1]{fontenc}
\usepackage{ae,aecompl}

\usepackage{graphicx}	
\usepackage{amsmath}	
\usepackage{amssymb}	
\usepackage{txfonts}

\title[PAH and galaxy evolution]{Modelling the evolution of PAH abundance in galaxies}

\author[S.-J. Rau, H. Hirashita, \& M. Murga]{
Shiau-Jie Rau$^{1,2}$\thanks{E-mail: cherrymow@gapp.nthu.edu.tw},
Hiroyuki Hirashita$^1$, and
Maria Murga$^3$
\\
$^{1}$Institute of Astronomy and Astrophysics, Academia Sinica,
Astronomy-Mathematics Building, AS/NTU,
No.\ 1, Sec.\ 4, Roosevelt Road, Taipei 10617, Taiwan \\
$^{2}$Department of Physics, National Tsing Hua University, Hsinchu 30043, Taiwan\\
$^3$Institute of Astronomy, Russian Academy of Sciences, Pyatnitskaya str.\ 48, Moscow 119017, Russia
}

\date{Accepted XXX. Received YYY; in original form ZZZ}

\pubyear{2019}

\begin{document}
\label{firstpage}
\pagerange{\pageref{firstpage}--\pageref{lastpage}}
\maketitle

\begin{abstract}
We investigate the evolution of polycyclic aromatic hydrocarbon (PAH)
abundance in a galaxy, which is a crucial step to understand the evolution of
bright emission features in the mid-infrared range.
We calculate the evolution of dust grain size distribution in a manner consistent with
the physical conditions of the interstellar medium by post-processing our previous
hydrodynamical simulation of an isolated disc galaxy. We also differentiate between aromatic and
non-aromatic grains for carbonaceous dust species and explicitly considered the
aromatization process.
As a consequence, our model explains the metallicity dependence of PAH abundances in
nearby galaxies well.
The PAH abundance increase is driven particularly by the interplay between
shattering and accretion (dust growth).
The fast aromatization guarantees that the small carbonaceous grains trace PAHs very well.
Since shattering and accretion are sensitive to the dust abundance, we predict that
the PAH-to-dust abundance ratio increases as the metallicity increases. This is
consistent with the observation data of nearby galaxies.
\end{abstract}


\section{Introduction}

Mid-infrared (MIR) spectral energy distributions 
of galaxies usually have prominent emission features.
Some of them are considered to be caused by carbonaceous species, which are likely to be
polycyclic aromatic hydrocarbons (PAHs) \citep{Leger1984,Allamandola1985,Li2012}.
There are some alternative candidates for the careers of these emission features
such as hydrogenated amorphous carbons (HAC; \citealt{Duley93}),
quenched carbonaceous composite \citep{Sakata83},
and mixed aromatic/aliphatic organic nanoparticles \citep{Kwok11}.
In this paper, we use PAHs to represent the materials responsible for the above
MIR emission features.\footnote{Detailed identification of these spectral features is not
important for this paper. As long as the emission features are attributed to small carbonaceous grains with
aromatic structures, the scenario of this paper holds.}
PAHs play a key role in the energy balance of the interstellar medium (ISM)
via absorbing ultraviolet (UV) photons and emitting in the MIR, and
the ionization balance via interaction with electrons and ions
\citep[e.g.][]{Tielens08}.
The luminosities of these emission bands are also used as an indicator of star formation in galaxies
\citep[e.g.][]{F04}.
Therefore, it is important to clarify the origin and evolution of PAHs in galaxies.


The luminosities of the MIR emission features show strong metallicity dependence
\citep[e.g.][]{E05,Ciesla14}.
The features are deficient in low-metallicity galaxies with active star formation such as
blue compact dwarf galaxies \citep[e.g.][]{Hunt10}.
Hard UV spectra \citep{Plante02}
and/or enhanced supernova (SN) activities \citep{O06} could play an important role in PAH destruction in low-metallicity environments.
	\citet{Madden06} also suggested the importance of strong UV radiation field for PAH destruction
	\citep[see also][]{Madden00,Wu06}.
	However, the indication that the size distribution of PAHs are shifted to small sizes
	in the Small Magellanic Cloud
	may be in tension with the destruction scenario, which should produce the opposite trend in PAH sizes
	\citep{Sandstrom12}.
It is also possible that PAH evolution is strongly linked to the dust evolution. Young
galaxies may not have sufficient time for asymptotic giant branch (AGB) stars to produce carbonaceous dust and PAHs and to supply them into the ISM \citep{G08}.
\citet{Bekki13} also explained the metallicity dependence of PAH-to-dust ratio by AGB stars.
However, \citet{Ch1992} calculated PAH formation yields in carbon-rich stellar outflows and
concluded that stellar outflows are not the main PAH formation sites.
We should also note that the metallicity does not necessarily reflect the age \citep[e.g.][]{Kunth2000}.
Thus, the lack of PAHs in low-metallicity galaxies may not be simply due to the age effect.
Since the metallicity dependence of the MIR features is also seen at
high redshift ($z\sim 2$; \citealt{Shivaei17}), it is important to clarify the physical reason for
the strong link between PAHs and metallicity in the context of galaxy evolution.

\citet [][hereafter S14]{Seok14} proposed a new idea to explain the metallicity dependence of PAH
abundance. They modelled the evolution of the PAH abundance in galaxies by
assuming that small grain production by shattering is the source of PAHs.
They also considered the destruction of PAHs by coagulation and SN shocks.
However, S14 did not directly solve grain size distribution in their model but applied the
\citet[][so-called MRN]{Mathis77} grain size distribution to estimate the formation and destruction
time-scales of PAHs.
Using these time-scales, they post-processed the evolution of the total dust amount calculated
by \citet{Asano13a}
and derived the evolution of the PAH abundance.
In this sense, their PAH abundance evolution was not self-consistently calculated with the evolution of grain size
distribution. Moreover, as shown by \citet{Hirashita12}, shattering is not the only source of small grains:
accretion drastically increases the small grain
abundance, while S14 implicitly assumed that accretion keeps the shape of the grain size
distribution.
Thus, we aim at a self-consistent treatment of grain size distribution and PAH formation,
although S14's basic idea that small-grain production leads to PAH formation still holds
in this paper.

Another point worth improving is an explicit treatment of aromatization, which
is a process of dehydrogenation
under the influence of UV radiation.
HAC grains have predominantly a disordered structure with aliphatic bonds between atoms.
These bonds, especially C--H bonds, can be dissociated under UV radiation,
and the number of hydrogen atoms in the HAC grains gradually decreases.
This leads to the formation of aromatic bonds and aromatic ring structures.
In this way, the structure becomes more ordered.
Thus, PAHs can result from photo-processing of HAC grains, as also shown by experiments~\citep{duley15}. The optical properties of the dehydrogenated HAC grains and PAHs are similar in the sense that
they have the same emission bands in the MIR range.
Since we are mainly interested in the abundance of aromatic emitters in this paper,
it is not important which species, PAHs or HAC, is responsible for them.
The essential points are (i) that there are aliphatic-dominated and aromatic-dominated species,
and (ii) that aromatization converts the former to the latter ones.
As mentioned above,
we call small grains responsible for the MIR emission features as PAHs, keeping in mind that they can be HAC grains as well.

The purpose of this paper is to model the evolution of PAH abundance in a consistent manner with the
evolution of grain size distribution. Since the grain size distribution is affected by
hydrodynamical evolution of the ISM, we solve the evolution of grain size distribution
based on hydrodynamical simulation results. We utilize the method developed by \citet [][hereafter HA19]{Hirashita19} for this purpose.
We further include the formation of aromatic species as a result of photo-processing of
(non-aromatic) HAC, rather than assuming
small carbonaceous grains as PAHs.
 

This paper is organized as follows. 
In Section \ref{model}, we describe our PAH evolution model.
In Section \ref{result}, we show the results and test them against observational data.
In Section \ref{discussion}, we compare our new model with previous results
(mainly S14), and describe future prospects.
In Section \ref{conclusion}, we give the conclusion of this paper.

\section{Model}\label{model}
We use the evolution model of grain size distribution in HA19, which was
developed based on \citet{Asano13}.
We post-process the hydrodynamical simulation of an isolated disc galaxy to derive
the evolutionary paths of grain size distribution in various conditions of the ISM.
In addition, we distinguish between aromatic and non-aromatic carbonaceous dust
and consider the aromatization reaction from the second to the first species.
For simplicity, we only treat the evolution of carbonaceous dust (see Section \ref{discussion}
for further discussions on this point).
In what follows, we explain our models.

\subsection{Hydrodynamic simulation}\label{subsec:hydro}

We adopt the same hydrodynamic simulation as used in HA19. We only give a brief
summary and refer the interested reader to HA19 for details.
We use the modified version of \textsc{gadget3-osaka} $N$-body/smoothed particle
hydrodynamics (SPH) code (\citealt{A17,Shimizu19}; the original version
of \textsc{gadget} was introduced by \citealt{Springel05}).
We adopt the initial condition used in the low-resolution model of AGORA
simulations \citep{Kim14,Kim16}. The following components are included in the initial condition:
halo, stellar disc, gas disc, and bulge. The gas mass resolution
is $8.6\times 10^4$ M$_{\sun}$.
The minimum gravitational softening length is 80~pc and the baryons are allowed
to collapse to 10 per cent of this value. The star formation is assumed to occur in
dense regions in a local free-fall time with an efficiency of 0.01.
Stellar feedback and metal enrichment are calculated by CELib \citep{Saitoh17}, which
includes both SNe and AGB stars.

The different evolution history of each SPH gas particle (hereafter, simply referred to as gas particle)
could lead to different grain size distributions and PAH abundances.
To concentrate on the typical galactic-disc region, we choose gas particles located at radii
$0.1<R<4$~kpc  and vertical height $ -0.3<z<0.3$ kpc in the cylindrical coordinate $(R,\, z)$.
Finally, we adopt 146 gas particles, which are sufficient to investigate the statistical
properties.

\subsection{Grain size distribution}\label{subsec:dustmodel}

We calculate the evolution of grain size distribution on each gas particle.
We assume grains to be spherical and compact,
so that $m=(4\upi /3)a^3s$, where $m$ is the grain mass, $a$ is the grain radius and $s$
is the material density of dust.
For simplicity, we assume that all the dust species are carbonaceous.
We separate the carbonaceous species
into aromatic and non-aromatic components, and concentrate only on
these two carbonaceous species in this paper.
We adopt the physical parameters (including the above grain density $s=2.24$\,g\,cm$^{-3}$;
see HA19 for the details in the material parameter dependence)
appropriate for graphite for simplicity. Indeed, as we show later, quick aromatization
would realize ordered structures rather than amorphous ones, which could justify
the usage of graphite material properties.
We neglect silicate to avoid complexity arising from
inter-species interactions (e.g.\ collisions between silicate and carbonaceous grains).
See Section \ref{subsec:caveats} for more discussions on this assumption.

We define the grain size distribution at time $t$, $n_i(a,\, t)$, such that $n_i(a,\, t)\,\mathrm{d}a$ is
the number density of grains with radius between $a$ and $a+\mathrm{d}a$.
The aromatic and non-aromatic species are specified by the index $i=\mathrm{ar}$
and nonar, respectively.
The total grain size distribution is indicated without an index as
$n(a,\, t)\equiv n_\mathrm{ar}(a,\, t)+n_\mathrm{nonar}(a,\, t)$.
The time evolution of the grain size distribution is driven by the
following processes: dust condensation in stellar ejecta, dust destruction by SN shocks,
grain disruption by shattering, dust growth by the accretion of gas-phase metals
in the ISM, and grain growth by coagulation.

The stellar dust production is calculated based on the increase of the metallicity
by assuming that a fraction $f_\mathrm{in}=0.1$
of newly injected metals from stars condense into dust.
We assume that stars produce large ($a\sim 0.1~\micron$) grains
(more precisely, we assume a lognormal grain size distribution centred at
$a=0.1~\micron$ for dust grains produced by stars).
All grains produced by stars are assumed to be non-aromatic (or we assume that
the grains become aromatic only after being processed in the ISM).

Shattering and coagulation occur in the collisions between grains whose velocities are
induced by interstellar turbulence \citep{Yan04}. High and low turbulence velocities in
the diffuse and dense ISM cause shattering and coagulation, respectively.
Shattering and coagulation are treated based on the so-called Smoluchowski equation.
We calculate shattering and coagulation separately for aromatic and non-aromatic
species for simplicity (this simplification does not affect our results significantly).
We assume that shattering only occurs in the
diffuse phase with $n_\mathrm{H}<1$~cm$^{-3}$ ($n_\mathrm{H}$ is the
hydrogen number density)
and that accretion and coagulation take
place in the dense phase with $n_\mathrm{H}>10$~cm$^{-3}$ and $T_\mathrm{gas} < 1000$ K
($T_\mathrm{gas}$ is the gas temperature).
Since the dense clouds where accretion and coagulation occur cannot be
spatially resolved in our simulation,
we assume that a mass fraction of $f_\mathrm{dense}=0.5$
in the dense phase
is condensed into dense clouds with hydrogen number density $10^3$ cm$^{-3}$ and
gas temperature 50 K on subgrid scales and
that coagulation occurs only in those dense clouds.

The evolution of grain size distribution by accretion and SN destruction is
governed by an advection equation in the $a$-space.
We only treat accretion in the subgrid dense gas (i.e.\ coagulation
and accretion take place in the same place).
We apply the
following procedure for accretion: we calculate the grain size distribution modified
by accretion in the next time-step, $n(a,\, t+\Delta t)$. Since accreted material form non-aromatic structures (since we assume that aromatization occurs in the
diffuse ISM; Section \ref{subsec:aroma}), we regard the increase of $n$ at
each $a$
as the increase of $n_\mathrm{nonar}$.
Therefore, the non-aromatic grain size distribution is derived by
$n_\mathrm{nonar}(a,\, t+\Delta t)=n(a,\, t+\Delta t)-n_\mathrm{ar}(a,\, t)$ if
$n_\mathrm{nonar}$ is positive. In fact, it could become negative because accretion
could decrease the dust abundance at the smallest $a$ bins. In this case, we assume
$n_\mathrm{nonar}=n$ and $n_\mathrm{ar}=0$.
For destruction, we count the SNe attacking the gas particle and assume
that each SN destroys the dust contained in the swept medium with a grain-size-dependent
efficiency given by HA19.


The total dust mass density $\rho_\mathrm{dust}(t)$ is calculated by
\begin{align}
\rho_\mathrm{dust}(t)=\int_0^\infty\frac{4}{3}\upi a^3s\, n(a,\, t)\,\mathrm{d}a.
\end{align}
The dust-to-gas ratio is estimated as
${\rho_\mathrm{d,tot}(t)}/\rho_\mathrm{gas}$,
where
$\rho_\mathrm{gas}=\mu m_\mathrm{H}n_\mathrm{H}$
($\mu =1.4$ is the gas mass per hydrogen, and $m_\mathrm{H}$ is the mass
of hydrogen atom).

In this paper, we neglect dust and PAH destructions by hard photons.
It is likely that photo-destruction occurs only in regions near to the intense UV sources.
Therefore, photo-destruction may not affect the
dust properties in the entire galaxy compared with the dust processing mechanisms included above.
Moreover, in the current simulation framework, a calculation of radiation field (or
radiation transfer), which is generally computationally expensive, is not implemented.
Photo-destruction also occurs selectively where the radiation field is hard; this means that
we also need to trace the spatial/time variation of the hardness of stellar SED.
Because of the local nature of photo-destruction and the technical difficulty,
we focus on the interstellar processing mechanisms described above.
Nevertheless, we include aromatization by the mean UV radiation field
in the next subsection to justify that
aromatic grains actually form on a short time-scale.
Thus, we also address the photo-destrcution by the mean UV field in Section \ref{subsec:caveats}.

\subsection{Aromatization}\label{subsec:aroma}

Owing to photo-processing, grains lose mainly their hydrogen atoms.
Herewith, the atomic structures change as mentioned above, and the material properties
change correspondingly. In particular, the band gap energy, $E_{\rm g}$, is related
to the number fraction of hydrogen atoms ($X_\mathrm{H}$) through
$E_{\rm g}=4.3 X_{\rm H}$ eV \citep{tamorwu90}.
When HAC grains are fully hydrogenated, $E_{\rm g}$ reaches the maximum value,  2.67~eV
\citep{jones13}. In contrast, $E_{\rm g}$ has the minimum value when the grains are
dehydrogenated. We assume that the minimum value of $E_{\rm g}$ is 0.1~eV.
The maximum and the minimum values of $E_{\rm g}$ correspond to
$X_{\rm H}=0.6$ and 0.02, respectively. We define the aromatization time as the time
necessary for dehydrogenation from the maximum value of $X_{\rm H}$ to the minimum one.

Based on \cite{jones14} and \cite{murga16a}, we use the following expression for the time
evolution of $X_\mathrm{H}$ for a grain:
\begin{equation}
\frac{\mathrm{d}X_{\rm H}}{\mathrm{d}t} = Y_{\rm diss}^{\rm CH} \sigma^{\rm CH}_{\rm loss}
\int_{10~{\rm eV}}^{13.6~{\rm eV}} Q_{\rm abs}(a, E) F(E) \,\mathrm{d}E,
\label{eq: arom}
\end{equation}
where $Y_{\rm diss}^{\rm CH}$ is the dissociation probability for an incident photon on the grain,
$\sigma^{\rm CH}_{\rm loss}(=10^{-19}~\mathrm{cm}^2)$ is the dissociation cross-section,
$F(E)$ is the photon flux with energy $E$,  and $Q_{\rm abs}$ is the absorption efficiency.
We adopt  $Y_{\rm diss}^{\rm CH}=1$ for $a<20$~\AA\ and
$Y_{\rm diss}^{\rm CH}=20$ \AA$/a$ for $a\geq 20$ \AA.
We use the absorption cross-sections described originally by \cite{jones12_3} and modified by
\citet{Murga19}.
The photon flux corresponds to the mean radiation field in the solar neighbourhood
given by \cite{mmp83}. This is valid for the purpose of examining the typical radiation field
in Milky-Way-type galaxies. Also, our simulated galaxy has a constant star formation rate similar to the
present Milky Way value ($\sim 1$ M$_{\sun}$ yr$^{-1}$).
Finally, we obtain the following fitting formula for the aromatization time:
\begin{align}
\frac{\tau_\mathrm{ar}}{\mathrm{yr}}=3\left(\frac{a}{\micron}\right)^{-2} + 6.6\times 10^{7}
\left(\frac{a}{\micron}\right) .\label{eq:arom_time}
\end{align}



Since aromatization is associated with UV irradiation, we assume
that it occurs in the diffuse ISM with
$0.1<n_\mathrm{H}<1$ cm$^{-3}$, where UV radiation can penetrate easily.
We define the aromatic fraction $f_\mathrm{ar}$ as
\begin{align}
f_\mathrm{ar}(a,\, t)\equiv{n_\mathrm{ar}(a,\, t)}/{n(a,\, t)},
\end{align}
and calculate its time evolution as
\begin{align}
\frac{\partial f_\mathrm{ar}}{\partial t} ={(1-f_\mathrm{ar})}/{\tau_\mathrm{ar}}.
\end{align}
Following S14, 
we assume the grain radii of PAHs to be between $a_\mathrm{PAH,min}=3\times 10^{-4}$ and
$a_\mathrm{PAH,max}=2\times 10^{-3}~\micron$, corresponding
to $\sim$20--3000 carbon atoms.
We calculate the mass density of PAHs, $\rho_\mathrm{PAH}$, as
\begin{align}
\rho_\mathrm{PAH}(t)=\int_{a_\mathrm{PAH,min}}^{a_\mathrm{PAH,max}}f_\mathrm{ar}(a,\, t)\,
\rho_\mathrm{dust}(a,\, t)\,\mathrm{d}a.
\end{align}

As clarified in the Introduction, we assume that small aromatized grains are the
careers of the MIR features. It is not necessary, however, to specify the optical
properties of large aromatized
grains in this paper. We expect that large aromatized grains do not show prominent
mid-infrared features. As shown by \citet{murga16b}, aromatic grains do not show
MIR features if grain radii are larger than $\sim 100$ \AA. Since we categorize
carbonaceous grains with regular ring-like structures as aromatized grains,
graphite could also be regarded as `aromatized' grains in our model.
Detailed specification of the material properties for large aromatized grains is left for
future work.

\section{Results}\label{result}

\subsection{Grain size distribution and aromatic fraction}
\begin{figure}
		\includegraphics[scale=0.43]{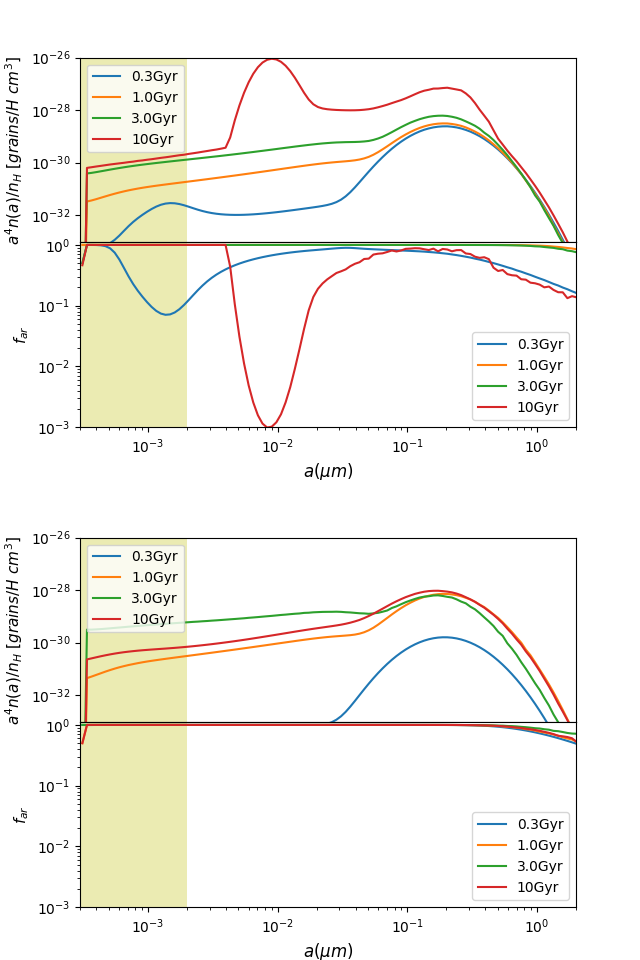}
		\caption{	
		Evolution of grain size distribution and aromatic fraction ($f_\mathrm{ar}$) for two of the gas particles.
		The upper and lower parts in each panel are the grain size distribution per hydrogen multiplied
		by $a^4$ (this quantity is proportional to the grain mass distribution per $\log a$),
		and the aromatic
		fraction ($f_\mathrm{ar}$), respectively. The different lines represent the different ages as shown in the legend.
		The shaded region shows the radius range of PAHs.
		}
	\label{fig:size}
\end{figure}

Since we already discussed the evolution of grain size distribution in HA19,
we only give a brief summary of it. 
The evolutionary trend of grain size distribution is  similar
among the gas particles; thus, we show only two  chosen cases
(Fig.\ \ref{fig:size}). To show the effect of dust growth by accretion, we
chose a case where the bump created by accretion is clearly seen
(upper panel of Fig.\ \ref{fig:size}).
At the earliest stage ($t\lesssim 0.3$~Gyr), stellar dust production dominates the dust abundance;
thus, the grain size distribution follows the lognormal
distribution of dust grains produced by stars.
At $t\gtrsim 0.3$~Gyr, shattering produces a tail towards small grain radii.
At $t\gtrsim 1$~Gyr, dust growth by accretion drastically increases the small grain abundance
because accretion is more efficient for smaller grains, which have larger surface-to-volume
ratios. Accretion often creates a prominent bump around $a\sim 0.01~\micron$
as observed in the upper panel of Fig.\ \ref{fig:size}.
At $t\gtrsim 3$~Gyr, coagulation converts small grains to large ones, contributing to
the increase of the large-grain abundance.

We also show the aromatic fraction ($f_\mathrm{ar}$) in the lower window of each panel.
We observe that
$f_\mathrm{ar}$ is almost 1 for
$a\lesssim 0.1~\micron$ at all ages. This is due to the short aromatization time-scale.
However, for the case shown in the upper panel in Fig.\ \ref{fig:size},
the aromatic fractions at $t=0.3$ and 3 Gyr have the minimum value at
$a\sim 10^{-3}~\mu\mathrm{m}$ and at $a\sim 10^{-2}~\mu\mathrm{m}$, respectively, where
the grain size distribution also has a peak.
This peak is caused by accretion as mentioned above.
Recall that accretion is assumed to contribute to the increase of non-aromatic grains
(Section \ref{subsec:dustmodel}).
As shown in equation (\ref{eq:arom_time}), the aromatization time-scale in the radius regime of PAHs is
$\sim 10^6$--$10^7$~yr.
The accretion time-scale can be as short as $10^6$ yr at $a\sim 10^{-3}~\micron$
at solar metallicity (HA19). This explains the temporary drop of aromatic fraction in the phase of
efficient accretion.
Since accretion occurs on a short time-scale, the aromatic fraction temporarily drops.
However, aromatization quickly occurs subsequently, and the aromatization time-scale is
still much shorter than the overall dust enrichment time-scale ($\sim $ chemical enrichment
time-scale). Thus, we do not observe
low aromatic fractions in most of the cases.
Except for the phase of rapid accretion in the dense ISM, we
can practically regard the production of small carbonaceous grains as the PAH formation.

\subsection{PAH abundance and metallicity}
\begin{figure}
	\includegraphics[scale=0.58]{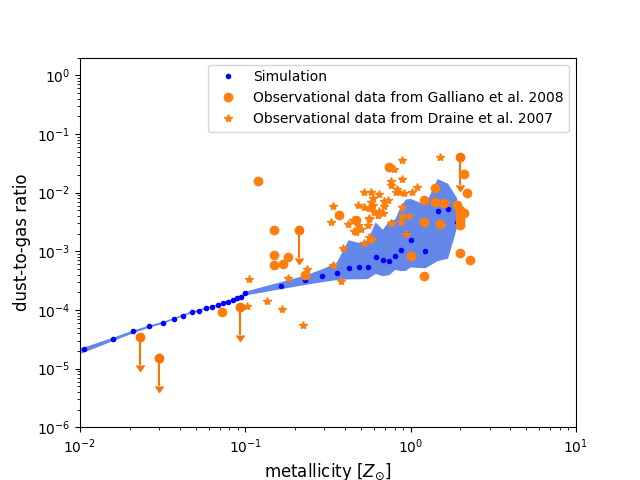}
	\caption{Relation between dust-to-gas ratio and metallicity. 
	The blue points are the median of our simulations in the metallicity bins with blue
	regions showing the area between the first and third quartiles. The orange points
	show the observational data for nearby galaxy samples taken from
	\citet{G08} and \citet{D07}. The typical error of dust abundance is smaller than a factor of 2.
	Note that there is also a factor 2 systematic uncertainty in solar metallicity, which causes a
	shift of the observational data in the horizontal direction.
	}
	\label{fig:dust_gas}
\end{figure}

\begin{figure}
		\includegraphics[scale=0.58]{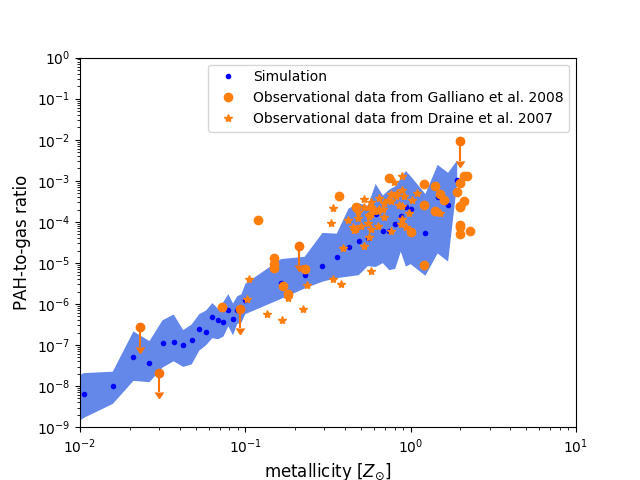}
		\caption{
		PAH-to-gas mass ratio as the function of metallicity.
			The blue points are the median of our simulations in the metallicity bins with blue
	regions showing the area between the first and third quartiles. The orange points
	show the data for the same observational samples as adopted
	in Fig.\ \ref{fig:dust_gas}. The typical error of
	the PAH abundance is a factor of 2.
	}
\label{fig:PAH_gas}
\end{figure}

\begin{figure}
		\includegraphics[scale=0.58]{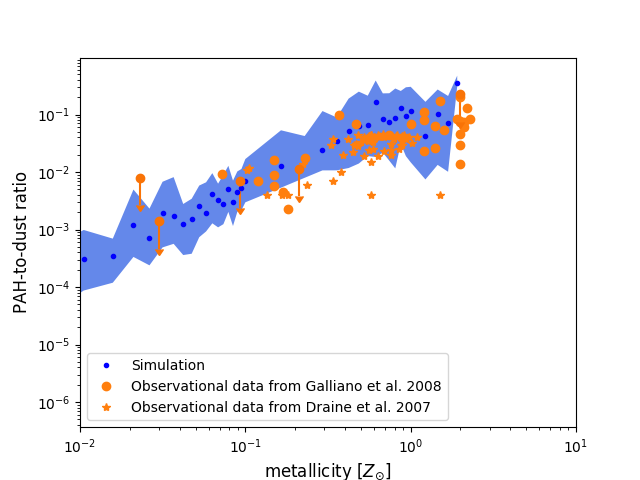}
		\caption{PAH-to-dust mass ratio as function of metallicity.
		The blue points are the medians in the metallicity bins with blue
	regions showing the area between the first and third quartiles for our simulation results.
	The orange points present
	the data for the same observational samples as adopted in Fig.\ \ref{fig:dust_gas}. 
	The typical error of the PAH abundance is a factor of 2.
	}
\label{fig:PAH_dust}
\end{figure}

Now we investigate the evolution of PAH abundance.
First, we examine the dust-to-gas ratio $\rho_\mathrm{dust}/\rho_\mathrm{gas}$
in Fig.~\ref{fig:dust_gas}
since the total abundance of dust, from which PAHs form in our model,
is a basic quantity for the interpretation of the PAH abundance.
For the indicators of PAH abundance,
we examine the PAH-to-gas ratio ($\rho_\mathrm{PAH}/\rho_\mathrm{gas}$) in Fig.~\ref{fig:PAH_gas} and
the PAH-to-dust ratio ($\rho_\mathrm{PAH}/\rho_\mathrm{dust}$) in Fig.~\ref{fig:PAH_dust}.
Metallicity is used as an indicator of evolutionary stage. Indeed, the relation between
dust-to-gas ratio and metallicity has been used to investigate the dust enrichment processes
\citep{Issa90,Schmidt93,Lisenfeld98,Dwek98,Remy14}.
We used all the snapshots (output every $10^7$ yr) for the
sampled gas particles to derive the following statistical properties.
We divided metallicity into 40 linear bins, and show the medians in these bins in the figures. 
The number of data points in each bin is around $10^3$.
The shaded regions represent the area between the first and third quartiles. 
We also compare the results with the data of nearby galaxies compiled by S14
(originally taken from \citealt{G08} and \citealt{D07}).
The observational metallicity data are based on the oxygen abundance.
Thus, we assume that the oxygen abundance fairly traces the total metallicity.
The solar oxygen abundance is uncertain, and we simply followed the solar
abundance values adopted in those original papers ($12+\log (\mathrm{O/H})_{\sun}=8.83$
and $12+\log (\mathrm{O/H})_{\sun}=8.69$ for
\citealt{G08} and \citealt{D07}, respectively), keeping in mind that there is
a factor 2 uncertainty in the solar metallicity.
The typical error of the PAH abundance is a factor of $\sim 2$.
Note that each of their data points represents an individual galaxy.
Here we regard our results on the gas particles as representing the evolution
of PAH abundance in galaxies with a variety of physical conditions.
In our post-processing procedure, it is not possible to derive the global quantities.
We expect that the global dust and PAH abundances lie near the medians or at least
within the shaded regions in the diagrams.


We observe in Fig.~\ref{fig:dust_gas} that the dust-to-gas ratio
increases with metallicity. This is a natural consequence of chemical enrichment and
metal accretion on dust.
Most of the observational data points are consistent with the simulation result.
The dust-to-gas ratio has very small dispersion at low metallicity ($Z\lesssim 0.1$ Z$_{\sun}$)
because the dust-to-metal ratio is always $\sim f_\mathrm{in}$ determined by the
condensation efficiency in stellar ejecta in our model.
A nonlinear increase and an expansion of dispersion are apparent at high
metallicity. This behaviour at high metallicity is caused by accretion (dust growth)
\citep{Dwek98,Hirashita99,Zhukovska08}.
We tend to underestimate the dust-to-gas ratio at subsolar metallicity
($\sim$0.2--0.5 Z$_{\sun}$), but this may be due to the effect of averaging.
Observational dust abundance is derived in a luminosity-weighted way,
which could be more biased to dust-rich regions associated with star-forming regions,
while we treat all the sampled gas particles equally.
However, we should emphasize that our model successfully covers a large part of
the observed dust-to-gas ratios at high metallicity.

Next, we examine the relation between PAH-to-gas ratio ($\rho_\mathrm{PAH}/\rho_\mathrm{gas}$)
and metallicity in
Fig.~\ref{fig:PAH_gas}.
Our model reproduces the observed relation.
The increase of PAH abundance coincides with that of dust-to-gas ratio
as expected, but
the increase of PAH abundance along the metallicity is steeper than that of dust-to-gas ratio.
For the purpose of showing this steep increase of PAH abundance,
we show the evolution of PAH-to-dust ratio
($\rho_\mathrm{PAH}/\rho_\mathrm{d,tot}$) in
Fig.~\ref{fig:PAH_dust}. Our calculation shows the increase of
PAH-to-dust ratio with metallicity. The increase is quantitatively consistent with the
observational data. The increase of PAH-to-dust ratio indicates that the PAH abundance
is more sensitive to the metallicity than the total dust abundance is.
Since PAHs originate from small grains in our model, shattering and accretion are important
for PAH formation. Accretion does not form PAHs directly, but it enhances the abundance of
small non-aromatic grains, which are subsequently converted to aromatic species.
Since the efficiencies of both accretion and shattering depend on the dust abundance
(and metallicity for accretion),
the increase of small grains has a nonlinear dependence on the metallicity
\citep{Hirashita15};
in other words, the increase of small grains is more sensitive to the metallicity than
that of the total dust abundance.
This is why the PAH-to-dust ratio increases with the metallicity.

\section{Discussion}\label{discussion}

\subsection{Comparison with the previous model}

S14 modelled the evolution of PAH abundance by post-processing the
dust-to-gas ratio calculated separately. They assumed shattering to be the
source of PAHs.
We focus on their model with the nearest star formation time-scale,
$\tau_\mathrm{SF}=5$ Gyr, to our simulated galaxy.
The evolution of PAH abundance in S14 shows a steep increase at
$Z\sim 0.2$ Z$_{\odot}$. This is a natural consequence of accretion, which drastically
raise the dust abundance at sub-solar metallicity.
S14 has a steeper increase of PAH abundance at $Z<0.2$ Z$_\odot$ than our results,
which is also due to their steeper increase of dust-to-gas ratio.
In a one-zone model like S14, the rapid increase of dust and PAH abundances by
accretion occurs at a certain metallicity, while in our spatially resolved modelling,
the rapid increase
does not occur coherently. As a consequence, the increase of PAH abundance becomes
milder in our model than in S14.

We calculated the grain size distribution, while S14 did not.
Accretion as well as shattering directly increases the small-grain
abundance. The rate of small-grain formation is treated in a consistent manner with the
evolution of grain size distribution in our model.
Although our results are qualitatively similar to S14's, the treatment of grain size distribution
is still essential since the rates of shattering and accretion strongly depends on the grain
size distribution.






Previous studies suggest other evolutionary pictures for PAH abundance.
Broadly, there are two major possibilities:
destruction by SNe and PAH supply by AGB stars.
\citet{O06} showed a prominent anticorrelation between the 7.7 $\micron$ feature strength,
which is an indicator of the PAH abundance, and the [Fe\,\textsc{ii}]/[Ne\,\textsc{ii}] line ratio,
which is a tracer of SN shocks.
They also found an anticorrelation between the [Fe \textsc{ii}]/[Ne \textsc{ii}] ratio and metallicity,
arguing that the strong destruction of PAHs by SN shocks could be the cause of the
deficiency of PAHs in low-metallicity galaxies.
We included shock destruction in our calculations, but found that shattering still continuously
supply small grains, which are quickly aromatized. In our model, shattering occurs everywhere in the
diffuse ISM because of the turbulent motion \citep[see also][]{Yan04,Hirashita09}. \citet{Jones96}
showed that grains are not only sputtered but also shattered in SN shocks.
Shattering in SN shocks could be as efficient as shattering in interstellar turbulence
\citep{Hirashita10}. Therefore, it cannot be simply concluded that the enhanced SN rate leads to
a depletion of PAHs.
For the other possibility (PAH supply by AGB stars),
\citet{G08} and \citet{Bekki13} assumed that the PAH abundance is governed by the ejection of carbonaceous dust
from AGB stars. In their models, the paucity of PAHs in low-metallicity galaxies is explained
by young ages (i.e.\ lack of AGB stars). 
Although our results do not exclude these possible explanation for the PAH--metallicity
correlation, we emphasize that our model naturally predicts it since the small-grain production is
sensitive to the metallicity.

\subsection{Caveats and future prospects}\label{subsec:caveats}

In our model, we only treated the carbonaceous species for simplicity.
In the Galactic environment, it is known that the interstellar dust contains as much silicate
as carbonaceous dust \citep[e.g.][]{Weingartner01}.
However, how these two species interact is not certain. They may evolve separately
or make compound species. Experimental
studies of dust growth could help to resolve this issue \citep{Rouille15}.
It is also suggested that carbonaceous materials could coat silicate grains \citep{Jones17}.
In this case, silicate grains act as a `host' of the growth of carbonaceous
grains in the sense that silicate grains accrete carbonaceous materials to form
grains whose optical properties eventually resemble pure carbonaceous grains.
If these coated grains are shattered (note also that there could be other mechanisms of
grain disruption; \citealt{Hoang19}), we expect that small carbonaceous dust is injected
into the ISM.
Therefore, we expect that our simplified treatment in this paper is still meaningful
in the presence of silicate.

The inverse reaction of aromatization, `aliphatization', could also happen.
This inverse process is also possible when the rate of dehydrogenation
is lower than that of hydrogen accretion on the grain surface.
As the C--C bonds become aliphatic after attachment of new atoms, the material structures
are transformed
from ordered to disordered ones.
Aliphatization is expected to occur in weak UV radiation field and high hydrogen density.
\cite{jones14} found that this process is efficient where the visual extinction is in the range
of 0.01--0.7 $\leq A_{\rm V} \leq 1.5$ (the lower limit depends on the incident radiation field intensity). 
On the other hand, we already included the formation of amorphous grains
through metal accretion in the dense ISM.
Aliphatization is also taken as an accretion process of hydrogen, so that
we can effectively include aliphatization into accretion in our framework.
We have shown that formed non-aromatic grains are aromatized quickly.
Thus, we expect that, even if we include the inverse reaction, our results
are not altered significantly because of the quick aromatization.
Moreover, since our results without aliphatization is consistent with the observed
PAH abundances, we could argue that aliphatization is not likely to play a dominant
role in determining the PAH abundance in nearby galaxies.

In this paper, we neglected photo-destruction, which is likely to be
effective only locally (Section \ref{subsec:dustmodel}). Nevertheless,
even the mean radiation field could destroy the smallest PAHs.
This destruction may systematically decrease the abundance of PAHs at all
metallicities compared with
our calculation results.
According to \cite{murga16a}, for the radiation field given by \cite{mmp83},
the destruction time-scale for PAHs smaller than $a\sim 5$ \AA\
is shorter than $\sim 10^9$ yr (dust enrichment time-scale).
However, even if all the PAHs
with $a\leq 5$ \AA\ are eliminated in our calculations, it decreases the PAH abundance
by at most 20 per cent. Therefore, the photo-destruction by the mean radiation field does not
affect our results significantly.

As a natural extension of our modelling, we could calculate the evolution of PAH
abundance and the hydrodynamical development of the ISM simultaneously.
This step serves to clarify the distribution of PAHs within a galaxy.
Indeed, there have already been some hydrodynamical simulations that include the
evolution of grain size distribution \citep{McKinnon18,Aoyama19}.
In addition, incorporating aromatization in the simulation framework
is necessary to complete this extension.

\section{Conclusion}\label{conclusion}
We post-process a previous hydrodynamical simulation of an isolated galaxy with
our evolution model of PAHs.
We calculate the evolution of PAH abundance in a consistent manner with
the evolution of grain size distribution. We also explicitly include aromatization.
In this model, the main PAH formation path is the interplay between
shattering and accretion and the subsequent aromatization.
We find that aromatization occurs quickly so that the formation of small carbonaceous
grains could be regarded as the production of PAHs.
Carbonaceous grains could be non-aromatic only in the phase of rapid dust growth by metal accretion
in the dense ISM, while they are aromatic in most of the galaxy evolution stages.

We find that the calculated relations between PAH abundance
(PAH-to-gas ratio and PAH-to-dust ratio) and metallicity are consistent with
the observation data of
nearby galaxies. In particular, the nonlinearity of shattering and
accretion in terms of the metallicity naturally explains the nonlinear increase of
the PAH abundance as a function of metallicity.
The dispersion of PAH abundances is interpreted as caused by a variety
in the physical conditions (gas density, temperature, etc.) of the ISM.
We conclude that the scenario that PAHs form as a result of small-grain production
by shattering and accretion with subsequent aromatization explains the observed strong metallicity
dependence of PAH abundance.

\section*{Acknowledgements}

We are grateful to the anonymous referee for useful comments.
We thank S. Aoyama for providing us with the hydrodynamic simulation data,
and J. Y. Seok for sending us the compiled data of nearby galaxies.
HH thanks the Ministry of Science and Technology for support through grant
MOST 105-2112-M-001-027-MY3 and MOST 107-2923-M-001-003-MY3.
MM acknowledges the support from the RFBR grant 18-52-52-006.




\begin{thebibliography}{99}
	
\bibitem[\protect\citeauthoryear{Allamandola et al.}{1985}]{Allamandola1985} 
Allamandola L.~J., Tielens A.~G.~G.~M., Barker J.~R., 1985, ApJ, 290, L25
\bibitem[\protect\citeauthoryear{Aoyama et al.}{2019}]{Aoyama19}
Aoyama S., Hirashita H., Nagamine K., 2019, \mnras, submitted (arXiv:1906.01917)
\bibitem[\protect\citeauthoryear{Aoyama et al.}{2017}]{A17} 
Aoyama S., Hou K.-C., Shimizu I., Hirashita H., Todoroki K., Choi J.-H., Nagamine K., 2017, MNRAS, 466, 105
\bibitem[\protect\citeauthoryear{Asano et al.}{2013a}]{Asano13a}
Asano R.S., Takeuchi T.T., Hirashita H., Inoue A.K., 2013a, Earth Planets Space, 65, 213
\bibitem[\protect\citeauthoryear{Asano et al.}{2013b}]{Asano13}
Asano R.S., Takeuchi T.T., Hirashita H., Nozawa T., 2013b, \mnras, 432, 637
\bibitem[\protect\citeauthoryear{Bekki}{2013}]{Bekki13}
Bekki K., 2013, \mnras, 432, 2298
\bibitem[\protect\citeauthoryear{Cherchneff et al.}{1992}]{Ch1992}
Cherchneff I., Barker J.~R., Tielens A.~G.~G.~M., 1992, ApJ, 401, 269
\bibitem[\protect\citeauthoryear{Ciesla et al.}{2014}]{Ciesla14}
Ciesla, L., et al., 2014, \aap, 565, A128
\bibitem[\protect\citeauthoryear{Draine et al.}{2007}]{D07}
Draine B. T. et al., 2007, \apj, 663, 866
\bibitem[\protect\citeauthoryear{Duley}{1993}]{Duley93}
Duley W. W., 1993, in S. Kwok, ed, ASP Conf.\ Ser.\ Vol.\ 41, Astronomical Infrared Spectroscopy:
Future Observational Directions, ASP, San Francisco, p.\ 241
\bibitem[\protect\citeauthoryear{Duley et al.}{2015}]{duley15}
{Duley} W.~W., {Zaidi} A., {Wesolowski} M.~J., {Kuzmin}, S., 2015, \mnras, 447, 1242
\bibitem[\protect\citeauthoryear{Dwek}{1998}]{Dwek98}
Dwek E., 1998, \apj, 501, 643
\bibitem[\protect\citeauthoryear{Engelbracht et al.}{2005}]{E05} 
Engelbracht C.~W., Gordon K.~D., Rieke G.~H., Werner M.~W., Dale D.~A., Latter W.~B., 2005, ApJ, 628, L29
\bibitem[\protect\citeauthoryear{F{\"o}rster Schreiber et al.}{2004}]{F04} 
F{\"o}rster Schreiber N.~M., Roussel H., Sauvage M., Charmandaris V., 2004, A\&A, 419, 501
\bibitem[\protect\citeauthoryear{Galliano et al.}{2008}]{G08}
Galliano F., Dwek E., Chanial P., 2008, \apj, 672, 214
\bibitem[\protect\citeauthoryear{Hirashita}{1999}]{Hirashita99}
Hirashita H., 1999, \apj, 510, L99
\bibitem[\protect\citeauthoryear{Hirashita}{2015}]{Hirashita15}
Hirashita H., 2015, \mnras, 447, 2937
\bibitem[\protect\citeauthoryear{Hirashita}{2012}]{Hirashita12}
Hirashita H., 2012, \mnras, 422, 1263
\bibitem[\protect\citeauthoryear{Hirashita \& Aoyama}{2019}]{Hirashita19}
Hirashita H., Aoyama S., 2019, \mnras, 482, 2555 (HA19)
\bibitem[\protect\citeauthoryear{Hirashita et al.}{2010}]{Hirashita10}
Hirashita H., Nozawa T., Yan H., Kozasa T., 2010, \mnras, 404, 1437
\bibitem[\protect\citeauthoryear{Hirashita \& Yan}{2009}]{Hirashita09}
Hirashita H., Yan H., 2009, MNRAS, 394, 1061
\bibitem[\protect\citeauthoryear{Hoang}{2019}]{Hoang19}
Hoang T., 2019, \apj, 876, 13
\bibitem[\protect\citeauthoryear{Hou et al.}{2017}]{Hou17}
Hou K.-C., Hirashita H., Nagamine K., Aoyama S., Shimizu I., 2017, MNRAS, 469, 870
\bibitem[\protect\citeauthoryear{Hunt et al.}{2010}]{Hunt10} 
Hunt L.~K., Thuan T.~X., Izotov Y.~I., Sauvage M., 2010, \apj, 712, 164
\bibitem[\protect\citeauthoryear{Issa, MacLaren, \& Wolfendale}{Issa et al.}{1990}]{Issa90}
Issa M. R., MacLaren I., Wolfendale A. W., 1990, \aap, 236, 237
\bibitem[\protect\citeauthoryear{Jones}{2012}]{jones12_3}
Jones A.-P., 2012, \aap, 542, A98
\bibitem[\protect\citeauthoryear{Jones et al.}{2017}]{Jones17}
Jones A.-P., K\"{o}hler M., Ysard N., Bocchio M., Verstraete L., 2017, \aap, 602, A46
\bibitem[\protect\citeauthoryear{Jones et al.}{2013}]{jones13}
{Jones} A.~P., {Fanciullo} L., {K{\"o}hler} M.,  {Verstraete} L., {Guillet} V., {Bocchio} M.,
{Ysard} N., 2013, \aap, 558, A62
\bibitem[\protect\citeauthoryear{Jones, Tielens, \& Hollenbach}{Jones et al.}{1996}]{Jones96}
Jones A. P., Tielens A. G. G. M., Hollenbach D. J., 1996, \apj, 469, 740
\bibitem[\protect\citeauthoryear{Jones et al.}{2014}]{jones14}
Jones, A.~P., {Ysard} N., {K{\"o}hler} M., {Fanciullo} L., {Bocchio} M., {Micelotta} E., {Verstraete} L.,
{Guillet} V., 2014, Faraday Discussions, 168, 313
\bibitem[\protect\citeauthoryear{Kim et al.}{2014}]{Kim14}
Kim J.-h. et al., 2014, \apjs, 210, 14
\bibitem[\protect\citeauthoryear{Kim et al.}{2016}]{Kim16}
Kim J.-h. et al., 2016, \apj, 833, 202
\bibitem[\protect\citeauthoryear{Kunth et al.}{2000}]{Kunth2000} 
Kunth D., {\"O}stlin G., 2000, A\&ARv, 10, 1
\bibitem[\protect\citeauthoryear{Kwok \& Zhang}{2011}]{Kwok11}
Kwok S., Zhang Y., 2011, \nat, 7371, 80
\bibitem[\protect\citeauthoryear{Leger \& Puget}{1984}]{Leger1984}
Leger A., Puget J. L. A\&A 137, L5-L8
\bibitem[\protect\citeauthoryear{Li \& Draine}{2012}]{Li2012} 
Li A., Draine B.~T., 2012, ApJ, 760, L35
\bibitem[\protect\citeauthoryear{Lisenfeld \& Ferrara}{1998}]{Lisenfeld98}
Lisenfeld U., Ferrara A., 1998, \apj, 496, 145 
\bibitem[\protect\citeauthoryear{Madden}{2000}]{Madden00}
Madden S. C., 2000, New Astronomy Reviews, 44, 249
\bibitem[\protect\citeauthoryear{Madden et al.}{2006}]{Madden06} 
Madden S.~C., et al.\ 2006, \aap, 446, 877
\bibitem[\protect\citeauthoryear{Mathis, Mezger, \& Panagia}{Mathis et al.}{1983}]{mmp83}
{Mathis} J.~S., {Mezger} P.~G., {Panagia} N., 1983, \aap, 128, 212
\bibitem[\protect\citeauthoryear{Mathis, Rumpl, Nordsieck}{Mathis et al.}{1977}]{Mathis77}
Mathis J.S., Rumpl W., Nordsieck K.H., 1977, \apj, 217, 425 
\bibitem[\protect\citeauthoryear{McKinnon et al.}{2018}]{McKinnon18}
McKinnon R., {Vogelsberger} M., {Torrey} P., {Marinacci} F., {Kannan} R., 2018, \mnras, 478, 2851
\bibitem[\protect\citeauthoryear{Murga, Khoperskov, \& Wiebe}{Murga et al.}{2016a}]{murga16a}
{Murga} M.~S., {Khoperskov} S.~A., {Wiebe} D.~S., 2016a, Astronomy Reports, 60, 233
\bibitem[\protect\citeauthoryear{Murga, Khoperskov, \& Wiebe}{Murga et al.}{2016b}]{murga16b}
{Murga} M.~S., {Khoperskov} S.~A., {Wiebe} D.~S., 2016b, Astronomy Reports, 60, 669
\bibitem[\protect\citeauthoryear{Murga, Wiebe, Sivkova, \&Akimkin}{Murga et al.}{2019}]{Murga19}
Murga, M.~S., Wiebe, D.~S., Sivkova, E.~E., et al.\ 2019, \mnras, 488, 965
\bibitem[\protect\citeauthoryear{O'Halloran et al.}{2006}]{O06}
O'Halloran B., Satyapal S., Dudik R. P., 2006, \apj, 641, 795
\bibitem[\protect\citeauthoryear{Rouill\'{e}, J\"{a}ger, \& Krasnokutski}{Rouill\'{e} et al.}{2015}]{Rouille15}
Rouill\'{e} G., J\"{a}ger C., Krasnokutski S. A., 2015, Faraday Discussions, 168, 449
\bibitem[\protect\citeauthoryear{Sandstrom et al.}{2012}]{Sandstrom12}
Sandstrom, K. M., et al., 2012, \apj, 744, 20
\bibitem[\protect\citeauthoryear{Saitoh}{2017}]{Saitoh17} 
Saitoh T. R., 2017, \aj, 153, 85
\bibitem[\protect\citeauthoryear{Sakata et al.}{1983}]{Sakata83}
Sakata A., Wada S., Tanabe T., Onaka T., 1983, \apj, 287, L51
\bibitem[\protect\citeauthoryear{Schmidt \& Boller}{1993}]{Schmidt93}
Schmidt K.-H., Boller T., 1993, Astronomische Nachrichten, 314, 361 
\bibitem[\protect\citeauthoryear{Seok et al.}{2014}]{Seok14}
Seok J.~Y., Hirashita H., Asano R.~S., 2014, MNRAS, 439, 2186 (S14)
\bibitem[\protect\citeauthoryear{Shimizu et al.}{2019}]{Shimizu19} 
Shimizu I., Todoroki K., Yajima H., Nagamine K., 2019, MNRAS, 484, 2632
\bibitem[\protect\citeauthoryear{Shivaei et al.}{2017}]{Shivaei17}
Shivaei I., et al., 2017, \apj, 837, 157
\bibitem[\protect\citeauthoryear{Springel et al.}{2005}]{Springel05} 
Springel V., 2005, MNRAS, 364, 1105
\bibitem[\protect\citeauthoryear{Plante \& Sauvage}{2002}]{Plante02} 
Plante S., Sauvage M., 2002, \aj, 124, 1995
\bibitem[\protect\citeauthoryear{R\'{e}my-Ruyer et al.}{2014}]{Remy14}
R\'{e}my-Ruyer A., et al., 2014, \aap, 563, A31
\bibitem[\protect\citeauthoryear{Tamor \& Wu}{1990}]{tamorwu90}
{Tamor} M.~A., {Wu} C.~H., 1990, Journal of Applied Physics, 67, 1007
\bibitem[\protect\citeauthoryear{Tielens et al.}{2008}]{Tielens08} 
Tielens A.~G.~G.~M., 2008, ARA \& A, 46, 289

\bibitem[\protect\citeauthoryear{Weingartner \& Draine}{2001}]{Weingartner01}
Weingartner J. C., Draine B.T., 2001, \apj, 548, 296
\bibitem[\protect\citeauthoryear{Wu et al.}{2006}]{Wu06}
Wu Y., Charmandaris V., Hao L., Brandl B. R., Bernard-Salas J.., Spoon H. W. W.,
Houck J. R. 2006, \apj, 639, 157
\bibitem[\protect\citeauthoryear{Yan, Lazarian, \& Draine}{Yan et al.}{2004}]{Yan04}
Yan H., Lazarian A., Draine B. T., 2004, \apj, 616, 895
\bibitem[\protect\citeauthoryear{Zhukovska, Gail, \& Trieloff}{Zhukovska et al.}{2008}]{Zhukovska08}
Zhukovska S., Gail H.-P., Trieloff M., 2008, \aap, 2008, 479, 453
\end{thebibliography}



\bsp	
\label{lastpage}
\end{document}